\documentclass[oldversion]{aa}
\usepackage{txfonts}
\usepackage{natbib}
\usepackage{graphicx}
\bibpunct{(}{)}{;}{a}{}{,} 
\newcommand{\mjup}{\,$M_\ensuremath{{\rm{Jup}}}$}

\newcommand{\teff}{\it{T}\rm$_{\rm{eff}}$}
\newcommand{\logg}{log\,\it{g}\rm}
\newcommand{\micron}{\,\ensuremath{\mu}\rm{m}}

\begin{document}
   \title{The highest resolution near-IR spectrum of the imaged \\
   planetary mass companion 2M1207\,b
   \thanks{Based on observations
          obtained at the Paranal Observatory, Chile for ESO program
          078.C-0800(B).}}

   \author{J.~Patience\inst{1}, R.~R.~King\inst{1}, R.~J.~De~Rosa\inst{1}
          \and
          C.~Marois\inst{2}
          }

   \institute{School of Physics, University of Exeter, Exeter, EX4
              4QL, UK\\
              \email{patience@astro.ex.ac.uk, rob@astro.ex.ac.uk, derosa@astro.ex.ac.uk}
         \and
             National Research Council Canada, Herzberg Institute of
              Astrophysics, Victoria, BC V9E 2E7, Canada\\
             \email{christian.marois@nrc-cnrc.gc.ca}
             }

   \date{Received Month dd, yyyy; accepted Month dd, yyyy}

 
  \abstract {Direct-imaging searches for planets reveal wide orbit
planets amenable to spectroscopy, and their atmospheres represent an
important  comparison to the irradiated atmospheres of Hot Jupiters. Using AO
integral field spectroscopy of  2M1207\,b, the shape of the continuum
emission over the $J$, $H$, and $K$ bands from the atmosphere of this  young,
planetary mass companion is measured in order to compare with atmospheric and
evolutionary models, and objects of similar temperature in young clusters and
the field. The 2M1207\,b spectrum has the highest spectral resolution
(R$\sim$300--1500) and largest wavelength coverage, including the first
$J$-band spectrum, for this benchmark object. The high signal-to-noise of the
data allow a clear identification of signatures of low surface gravity, and
comparison with a grid of AMES-Dusty models reveals a best-fit effective
temperature of \teff=1600\,K with a preferred surface gravity of \logg=4.5.
The $J$-band flux is depressed relative to nearly all L-type objects, and the
detailed shape of the absorption features across the $H$-band exhibit
differences from the model predictions. The possible origins of 2M1207\,b and
its low luminosity are examined with the new data and analysis which
suggest that extinction from a disk with large grains is a viable scenario
and is preferred over scatttering off an optically thick disk. The 2M1207\,b
spectrum presents an important comparison for the  types of features which
may be present in upcoming spectra of the atmospheres of planets imaged in
orbit around stellar primaries.}



   \titlerunning{The highest resolution spectrum of 2M1207\,b}

   \maketitle
%

\section{Introduction}

Over 400 extrasolar planets are currently known\footnote{www.exoplanet.eu},
with most systems detected indirectly in orbits
closer than that of Jupiter. For indirectly detected planets, the rare
combination of a favourable geometry resulting in a transit and a very bright
host star for high signal-to-noise data is required to obtain a planetary
spectrum either in transmission \citep{c02} or emission \citep{ch05,d05}.
Although only two transiting planets have been investigated spectroscopically,
transmission studies have shown the possibility of methane absorption
\citep{sw08} or Rayleigh scattering \citep{s08} and a haze layer \citep{p08},
and emission studies have observed the presence \citep{g08} and absence
\citep{r07} of water band absorption in different systems.

The atmospheres of directly-imaged planets represent a complementary
case study to the transiting planet spectroscopy, uncontaminated by the
physical effects of extreme proximity to the host star. In some cases,
spectroscopy of directly imaged planets is feasible. The first confirmed
directly-imaged planets orbiting normal stars \citep{k08,m08} and the first
imaged planetary mass companion \citep{c04} present the opportunity for
spectroscopic analysis of young planet atmospheres. For technical reasons,
targets observed in imaging planet searches are typically  younger than stars
searched for planets indirectly, and this selection criterion enables an
investigation of the early evolution of planetary atmospheres. The first
spectrum of HR\,8799\,c in the $L$-band suggests differences in the continuum
shape from theoretical models \citep{j10}, possibly due to
non-equilibrium CO/CH$_4$ chemistry and indicating the importance of
measuring the atmosphere over an even larger wavelength range. This paper
presents the results of the highest resolution $J$, $H$, $K$ spectrum of
2M1207\,b which should provide a high signal-to-noise comparison spectrum to
planets imaged in orbit around stellar primaries.

\begin{table*}
\caption{SINFONI observations}             
\label{table:3}      
\centering          
\begin{tabular}{l c c c c l l l l l}     
\hline\hline       
                      
Source & Grating & Scale & NDIT & DIT & $\#$exp. & Standard
(HIP) & \teff\ (K) &
Dates & Project ID \\ 
\hline  

   2M1207A+b   & $H$+$K$ & 100 & 1 & 300\,s   & 24 & 54930, 62566 &
                29000, 15000 & 28\,Jan\,07, 07\,Feb\,07  & 078.C-0800(B)    \\
               & $J$   & 100 & 1 & 600\,s   & 6 & 59951 &  16400 & 22\,Feb\,07 & 078.C-0800(B) \\  

\hline                  
\end{tabular}
\end{table*}


\section{Target}

The target for this study is one of the lowest mass companions imaged to date
\citep{c04,c05}. 2M1207\,b is confirmed as a companion based on multi-epoch
imaging \citep{s06} and, considering the distance of 52.4$\pm$1.1\,pc
\citep{d08}, orbits the host brown dwarf at a projected separation of 46\,AU, somewhat
wider than the Sun-Neptune distance. The companion has apparent magnitudes of
$J$=20.0$\pm$0.2 \citep{m07}, $H$=18.09$\pm$0.21, and
$K_{\rm{S}}$=16.93$\pm$0.11 \citep{c04}. Combining the system age and
theoretical cooling tracks \citep{c00,b02,b03} with the target photometry
\citep{c05} indicates a mass of $\sim$5\mjup, whereas with the
spectroscopically estimated effectve temperature, a mass of 6--10\mjup\ is
predicted \citep{m07}.  The masses depend critically on the system age, estimated to be
8\,Myr for 2M1207 from its membership in the TW Hydra association \citep{g02};
the unusually well-determined age of this low mass companion eliminates some
degeneracies in model comparisons. Given the large difference in mass ratio of
the 2M1207 system compared to the HR 8799 and Fomalhaut systems, their
formation mechanisms may be different, with 2M1207 representing the tail of
the binary distribution and the A-star planets examples that formed in a disk
via core accretion \citep[e.g.,][]{p96} or gravitational instability
\citep[e.g.,][]{b97}. Despite the possibility of different origins,
2M1207\,b is a benchmark object for the study of young planetary
atmospheres. 

%
%

\section{Observations}

The observations were taken with SINFONI, an AO-equipped integral field
spectrograph  mounted at the Cassegrain focus of the VLT \citep{b04,e03,t98}
in program 078.C-0800(B) (PI: N. Thatte). Details of the observations
are given in Table\,\ref{table:3}. The primary object was used for the AO
correction. Of the four gratings available in SINFONI, the $J$ (R$\sim$2000)
and $H$+$K$ (R$\sim$1500) were used, which yields higher spectral resolution
than previous results  \citep{c04,m07} and the first $J$-band spectrum. The
100 mas spatial scale was  employed as it is fine enough to separate the
pair, but coarse enough to concentrate the flux of the faint companion. The
field-of-view at this scale is 3$\farcs$0 $\times$ 3$\farcs$0 and included
both components.  

The observing sequence included a set of exposures on the target at several
offset positions on the array. Observations of an early type star followed
each target sequence to correct for telluric features and the instrument
response. Additional calibrations were obtained to measure the wavelength
scale, the detector dark current, distortion, and quantum efficiency
variations.  

%
%

\section{Data Analysis}
\label{data_analysis}

The data were reduced with the Gasgano \citep{g07} implementation of the
SINFONI data reduction pipeline \citep{m09,d07}. With a set of detector
calibration files, corrections for the dark current, linearity response, and
bad pixels were measured and applied to all data files. Two effects of the
instrument optics were also corrected -- the distortion of the pixel scale
across the field and the non-linear dispersion of the gratings.  After
the detector and optics processing was completed, the array data were
converted into data cubes consisting of a stack of images, each covering a
slice of wavelength. Across the entire wavelength range, a stack of images
were constructed, composed of $\sim$700--2000 wavelength slices, depending on
the band. Binning was required for the $J$-band data.

\begin{figure}
\centering
\includegraphics[width=\columnwidth]{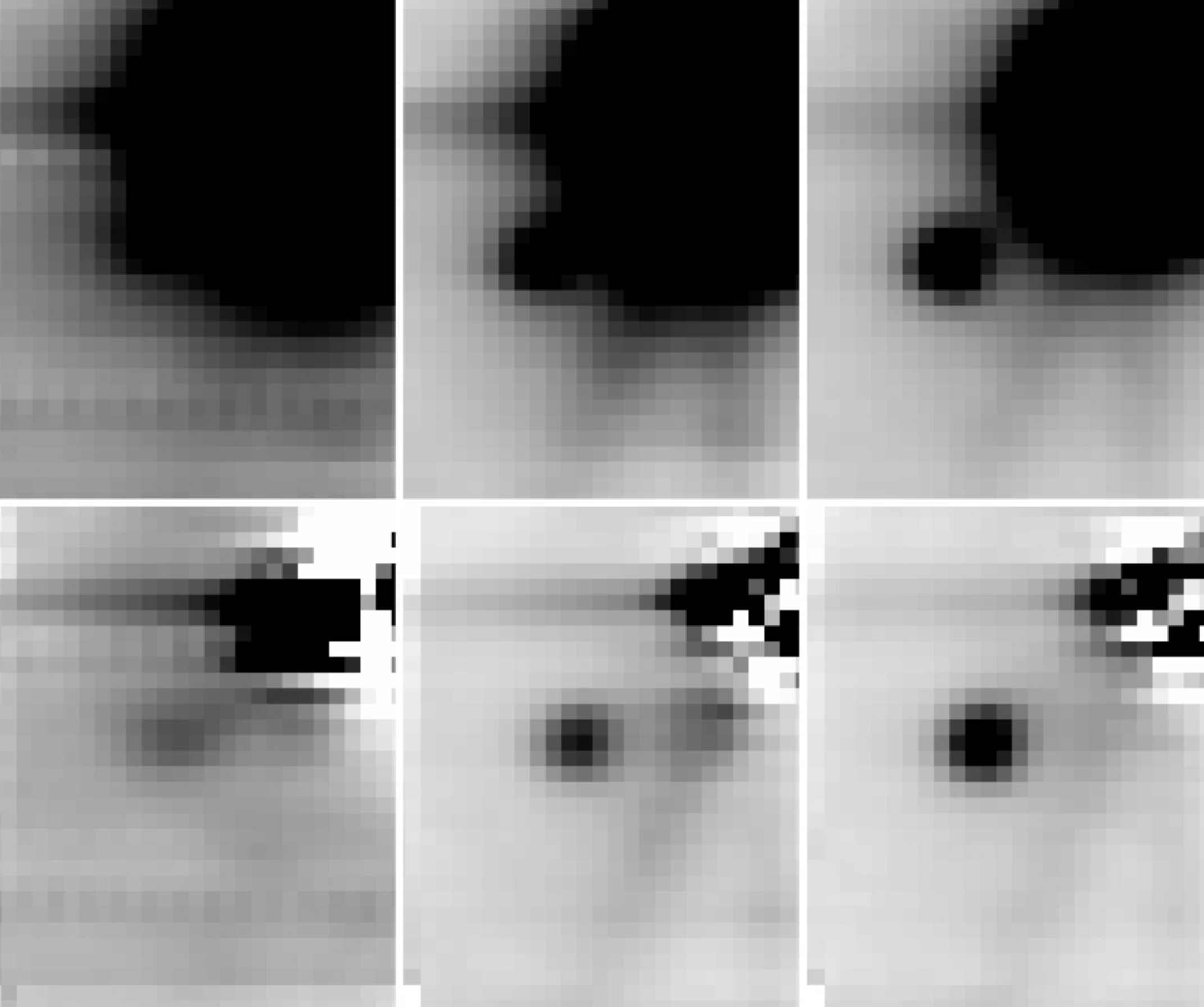}

   \caption{The $J$,$H$,$K$ wavelength collapsed images before and
   after the alignment and subtraction process. For the $H-$ and $K-$
   band portions of the data, the companion is very distinct; 
   although the $J-$band image is lower signal-to-noise,
   the companion is still clearly evident, enabling the spectrum to
   be measured in the binned data cube.}

      \label{Images}
\end{figure}

The final steps applied to the data cube involved sky subtraction,
combining the offset positions, radial subtraction of the host object, and
tracing the path of the companion spectrum through the data cube. The targets
were observed at different positions on the array in an ABBA pattern sampling
two areas on the array. The cubes with the target in one portion of the array
were averaged and then used as the sky to subtract off the individual cubes
at the opposite position. Next, the individual sky-subtracted cubes were
combined, centroiding on the primary. The individual wavelength planes were
shifted and aligned on the primary such that the final path of the companion
through the data cube was a straight trajectory. Due to the small angular
separation between the primary and secondary, the halo of the primary was fit
with a radial profile and subtracted from each wavelength plane. Images at
$J$, $H$, and $K$, collapsed in wavelength over the entire band, before and
after the subtraction process are shown in  Fig.\,\ref{Images}. The measured
{\it FWHM} of the target was stable at $\sim$3.5 pixels. Following the radial
subtraction, the companion spectrum was constructed by extracting the flux in
an aperture of 4.5 pixel radius that encompasses the majority of the flux in
a consistent manner. Given the stability of the primary FWHM throughout each
wavelength band, variations of the Strehl ratio across a band seem minimal
and should not systematically bias the extracted spectrum.

   \begin{figure*}
   \centering
   \includegraphics[width=\textwidth]{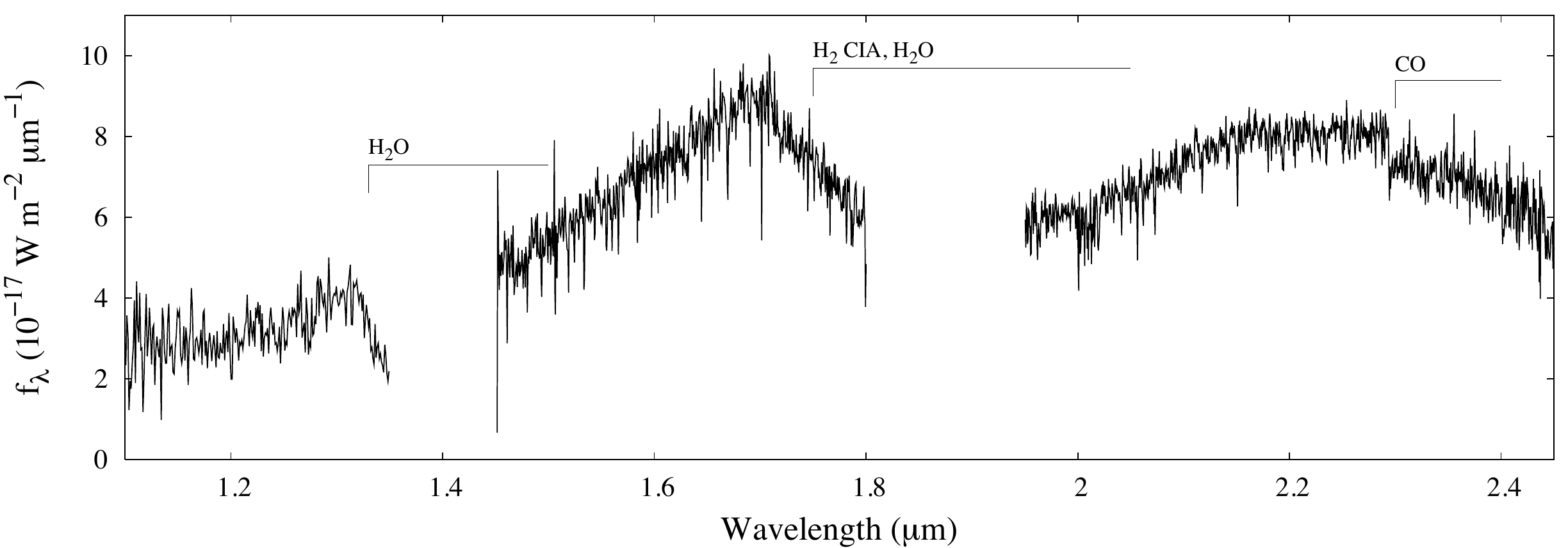}

      \caption{The spectrum of 2M1207\,b covering the $J$, $H$, and $K$
passbands (black lines) flux calibrated using the apparent magnitudes of
\citet{c04} and \citet{m07}. Key features are indicated.}

         \label{2M1207b-spectrum}
   \end{figure*}

Correction for telluric absorption and the detector response was accomplished
by dividing the observed target spectra by those of early-type standard stars
with interpolation across intrinsic features. The shape of the standard was
then removed by multiplication by a black-body at the appropriate effective
temperature (see Table\,\ref{table:3}). The extracted $J$ and $HK$
spectra of 2M1207\,b were flux calibrated using the $J$ and $K_{\rm{S}}$-band
photometry of \citet{m07} and \citet{c04}, respectively. The expected ratio
of detected photons was calculated by convolving the observed spectrum and
that of Vega \citep{Mountain:1985,Hayes:1985} with the 2MASS responses curves
\citep{Carpenter:2001}, assuming Vega to have zero magnitude in both filters.

The flux calibrated near-IR spectrum of 2M1207\,b is shown in
Fig.\,\ref{2M1207b-spectrum}. Over the $J$-band the signal-to-noise
ratio is $\sim$7 per resolution element, increasing to
$\sim$13 across the $H$-band and $\sim$24 over the $K$-band. The
depression of $J$ due to atmospheric 
dust absorption, the two prominent water absorption bands between the near-IR
passbands, and the CO absorption at the red end of the $K$-band are
clearly evident.

To estimate the effective temperature and surface gravity of 2M1207\,b, we
compared the flux calibrated $JHK$ spectrum to the AMES-Dusty model grid
\citep{a01} with \teff=500--4000\,K and \logg=3.5--6.0.  The model
spectra were smoothed to the same resolution as the 
observations and interpolated to the same dispersion. Using the same
least-squares statistic as \citet{m07} and allowing the model flux to be
scaled to find the best global fit, we identified the best fitting models.
Calibrating the model spectra to the flux as would be observed from Earth
requires a scale factor which depends on the radius of the object and its
distance. Since we have flux calibrated the observed spectrum, we can extract
an estimate of the radius implied for the closest match between each model and
the observations.

%
%

\section{Results and Discussion}
\label{results}

\begin{figure}
   \centering
   \includegraphics[width=\columnwidth]{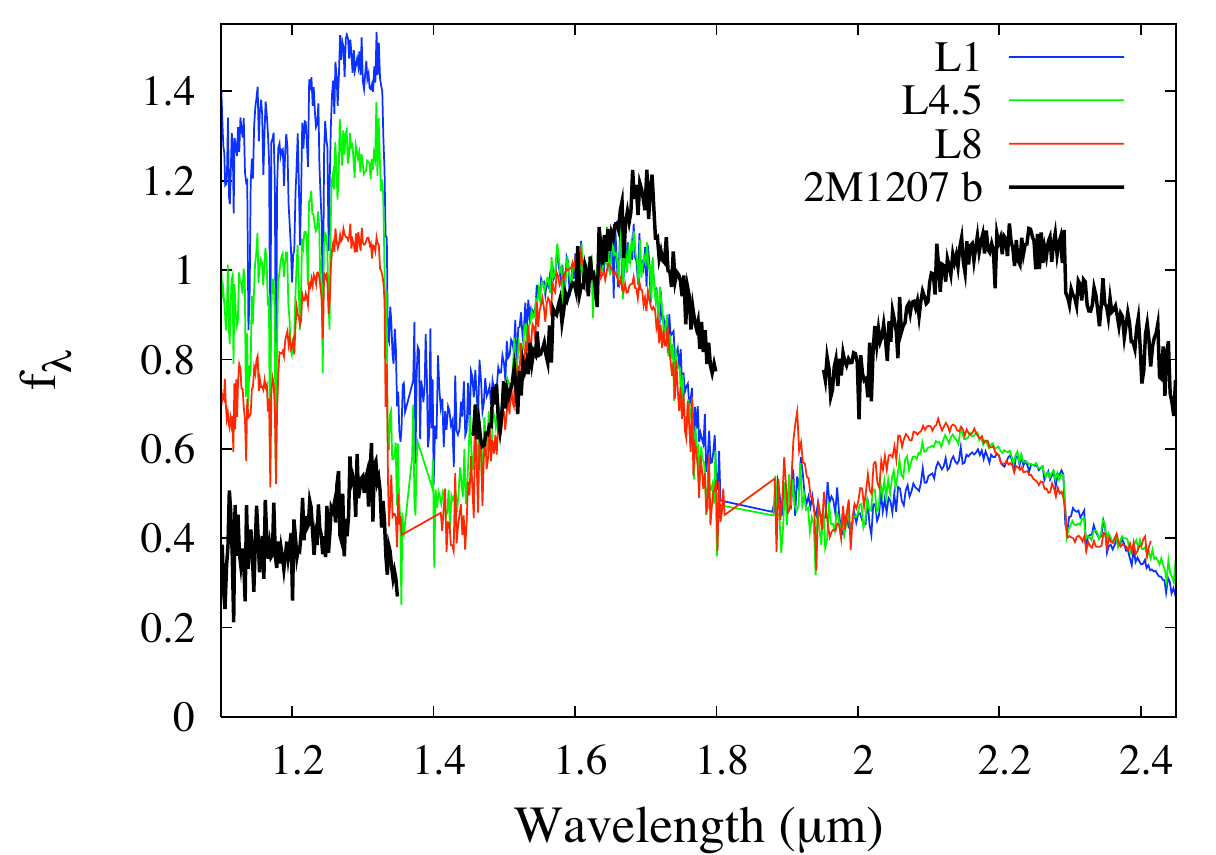}

      \caption{The $J$,$H$,$K$ spectrum of 2M1207\,b (black line) and several
comparison spectra of field L dwarfs with spectral types L1 (blue), L4.5
(green), and L8 (red). All spectra are normalised over the 1.60-1.65\micron\
range.}

         \label{FieldComp}
   \end{figure}

The larger coverage in wavelength and higher signal-to-noise enables an
investigation of the atmosphere and origin of this young, planetary mass
companion. The flux calibrated spectrum of 2M1207\,b is compared with previous
observations of field and young low mass objects and with theoretical model
atmospheres. The overall shape is significantly different to field L dwarfs,
with a marked depression in the $J$-band as shown in Fig.\,\ref{FieldComp}.
The $H$-band continuum exhibits a distinct triangular shape unlike the
older, higher surface gravity
objects, and the $K$-band peaks at $\sim$2.3\micron\ \citep[as suggested
by][]{m07} rather than at $\sim$2.15\micron\ as with field L dwarfs, and the
region associated with the peak is more clearly defined in these higher
resolution data.

\begin{figure}
   \centering
   \includegraphics[width=\columnwidth]{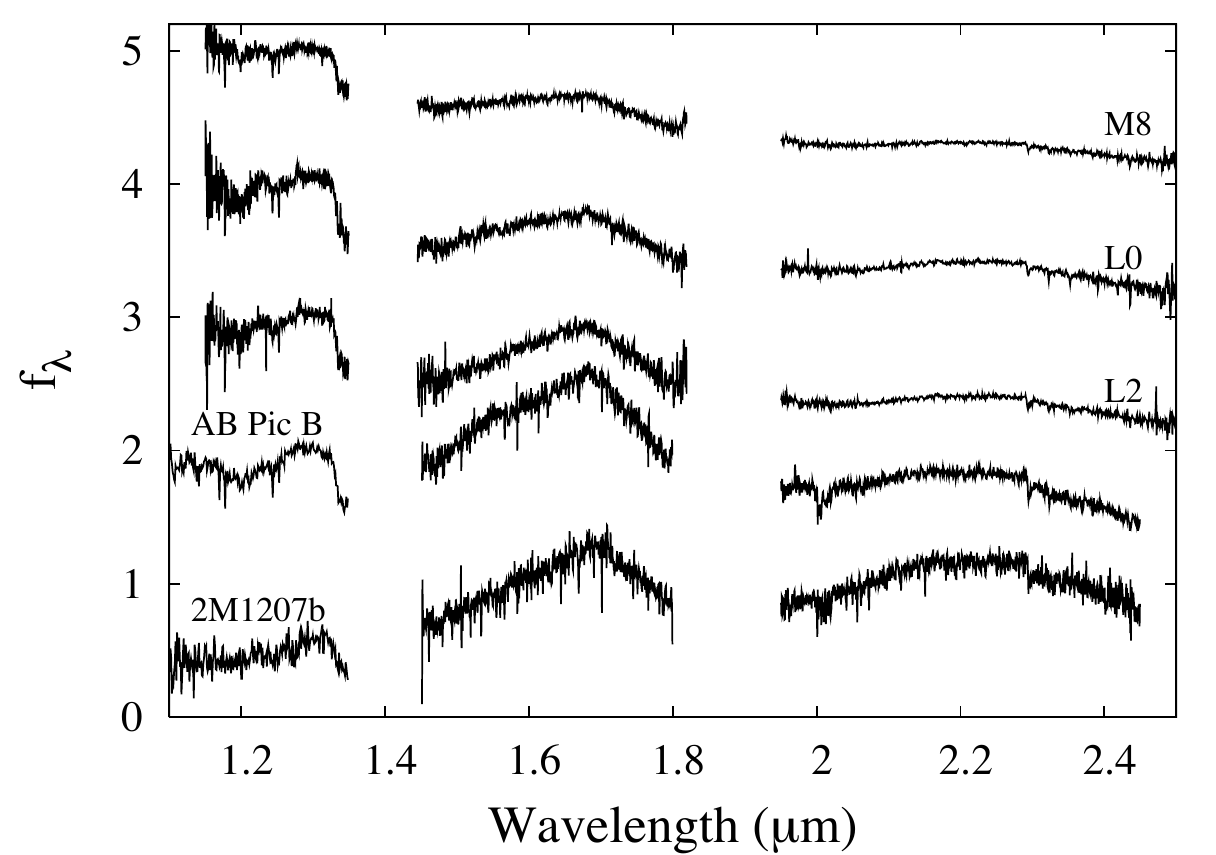}

      \caption{The $J$,$H$,$K$ spectrum of 2M1207\,b (bottom) and several
              comparison spectra of young (5--30Myr) objects in
              Upper Sco of spectral
              types (from the top): M8, L0, L2, and $\sim$L0--L1 (AB Pic b).
              The spectra have been flux calibrated as in Figure 1, and then offset by an integer for clarity. }

         \label{YoungComp}
   \end{figure}

The spectrum of 2M1207\,b is unlike field L-dwarfs spanning a range of
spectral types, but the young age of the TW Hydra association (8 Myr)
complicates the comparison. In Fig.\,\ref{YoungComp}, the spectrum of
2M1207\,b is plotted with several spectra of low mass members of Upper Sco
\citep{l08}, with spectral types ranging from M8--L2, and with the borderline
brown dwarf/planet companion AB Pic\,b \citep{b10}. Even amongst young, low-mass objects
in the 5\,Myr Upper Sco region, the 2M1207\,b spectrum appears distinct in the
shape and relative fluxes of the continuum across the different bands. No
Upper Sco member has a lower $J$-band relative to $H$-band and the triangular
shape of the $H$-band is only beginning to develop by the L2 spectral type.
The spectral sequence shown in Fig.\,\ref{YoungComp} suggests that 2M1207\,b
is one of the coolest young companions imaged. The most analogous object to
2M1207\,b is the companion AB Pic\,b which also shows some of the key features
present in  2M1207\,b. Since the $J$-band data required binning, the spectral
resolution is not yet high enough to identify and measure absorption lines,
though future observations with more signal-to-noise could investigate the
depths of the \ion{Na}{I} and \ion{K}{I} lines.

A series of spectral indices based on ratios of the flux of different regions
of the continuum have been developed to characterise the spectral type of cool
objects \citep{m03,s04,a07}. We measured these for 2M1207\,b and summarise the
results in Table\,\ref{spectral_indices} for the \citet{a07} H$_2$O and
gravity sensitive indices, the \citet{s04} H$_2$O and FeH indices, and the
\citet{m03} H$_2$O indices. These suggest a spectral type in the broad range
M8.5--L4, while the gravity sensitive index is consistent with that found for
other young brown dwarfs. The discrepancy in
inferred spectral types is not surprising given the unusual overall shape of
the spectrum and highlights the importance of obtaining larger wavelength
coverage of flux calibrated spectra in assigning spectral types and their
uncertainties.

\begin{figure*}
   \centering
   \includegraphics[width=\textwidth]{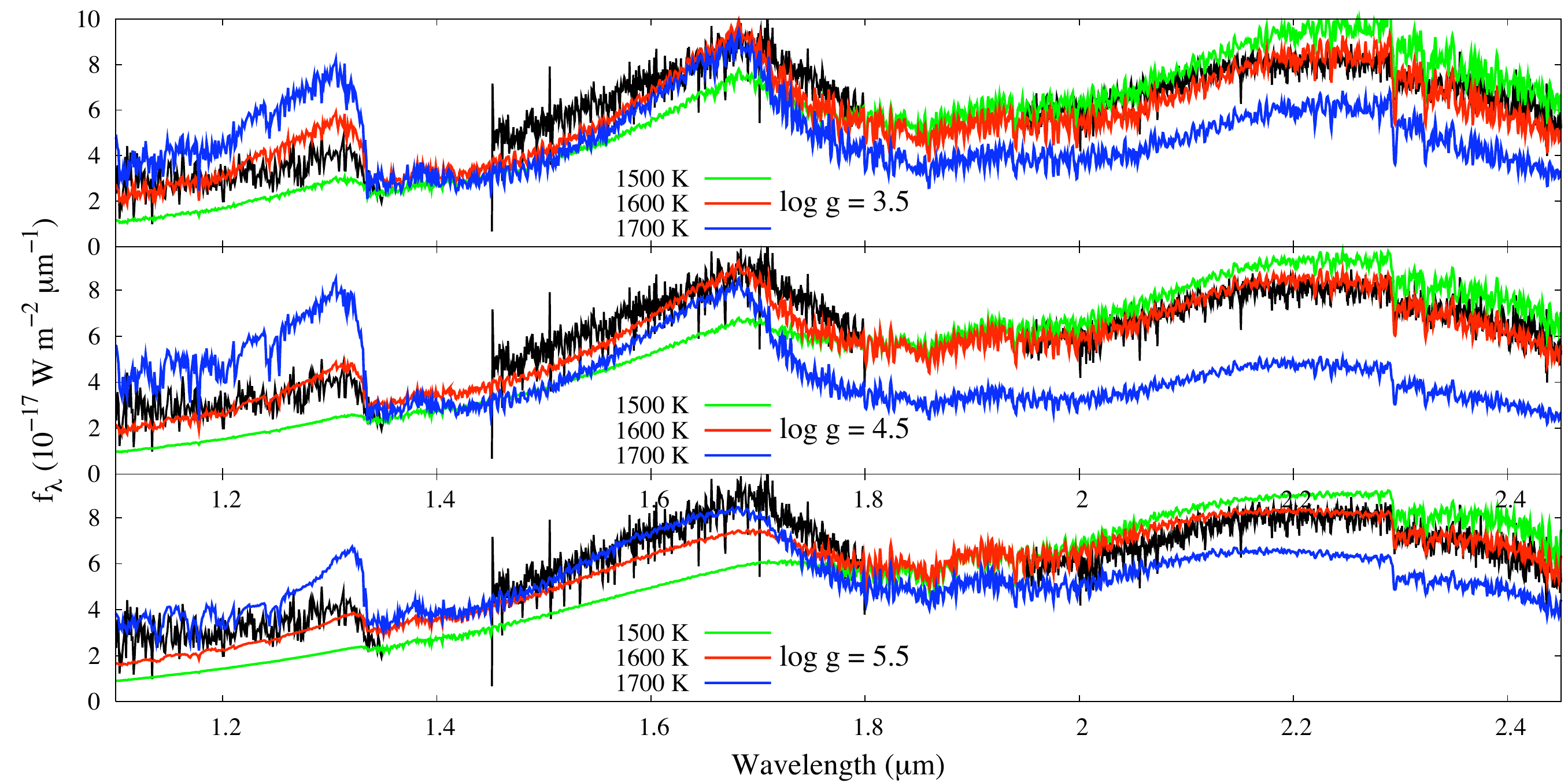}

      \caption{The spectrum of 2M1207\,b covering the $J$, $H$, and $K$
passbands (black lines). {\it Middle panel:} the best-fit DUSTY model with
\teff=1600 K and \logg=4.5 in red along with the 1500\,K (green line) and
1700\,K (blue line) \logg=4.5 models. {\it Top and bottom panels:} DUSTY
models of the same effective temperature as the middle panel, but with
surface gravity of \logg=3.5 and \logg=5.5, respectively.\rm The
observed spectrum has been flux calibrated using the apparent magnitudes of
\citet{c04} and \citet{m07}.  While the DUSTY models reproduce the
general morphology, the flux scaling necessary implies a radius of
$\sim$0.052\,$R_{\sun}$ at a distance of 53\,pc, significantly smaller
than any predicted radius of an ultracool dwarf. Discussion of this
unphysical radius, or equivalently low luminosity is given in Sect.\,\ref{results}.}

         \label{2M1207b-spectra}
   \end{figure*}

The flux calibrated 2M1207\,b spectrum with R$\sim$300 at $J$ and R$\sim$1500
at $HK$ is shown in Fig.\,\ref{2M1207b-spectra} along with the best fit to
the model DUSTY spectrum \citep{a01}. Available models with temperatures
higher and lower by 100\,K are also plotted in Fig.\,\ref{2M1207b-spectra}.
The best fit, estimated by a least-squares fit to the grid of DUSTY models,
and confirmed by visual examination is shown in the middle panel of
Fig.\,\ref{2M1207b-spectra} with \teff=1600 K and \logg=4.5. The uncertainty
in temperature is clearly less than $\pm$100\,K, but the ability to
discriminate surface gravity variations with the broad morphology and flux
level of the spectrum is less than for effective temperature.  Consequently,
\logg=4.5 is only marginally favoured over \logg=3.5 or 5.5 due to the
uncertainty on the flux level of the $J$-band spectrum resulting from the
photometry uncertainty of 0.2$^{\rm{m}}$. While the location of the peak in
the $H$-band is consistent with the models, the observed slopes on either
side are not as steep as the model predicts, possibly due to difficulties
modelling the H$_2$ collisionally-induced absorption. If the precision
of the $J$-band photometry could be improved, the model variations would
allow surface gravity to be constrained to $\sim$0.5 dex, important to test
one formation model.  These physical parameters are consistent  with those
measured previously \citep{m07}, though the inclusion of the $J$-band and the
higher signal-to-noise of these data makes the fitting more
secure. 

We did not compare the data to the COND models since they  do not
adequately reproduce the previous lower resolution $H$- and $K$-band spectrum
\citep{m07}. In addition to the treatment of dust condensation and settling,
the inclusion of non-equilibrium CO/CH$_4$ chemistry can significantly alter
the shape of the continuum emission \citep{f08}, particularly in the $K$- and
$L$-bands. A suitable parameterisation of the eddy diffusion coefficient for
such models may reproduce the shape of the observed spectrum, however, the
flux level for the \teff=1400\,K (the hottest shown), K$_{zz}$=4
non-equilibrium atmosphere shown in \citet{f08} is still significantly higher
than 2M1207\,b; for the range of K$_{zz}$, \logg, and metallicity
presented for the  \teff=1400\,K case, the $K$-band magnitude varies by up
to 0.7\,mag. 

\begin{table}
\begin{minipage}[t]{\columnwidth}
\caption{Spectral indices from Allers et al. (2007), Slesnick et al.
(2004), and McLean et al. (2003).}
\label{spectral_indices}
\centering
\renewcommand{\footnoterule}{}  
\begin{tabular}{llllll}     
\hline\hline       
                      
Index & H$_2$O   & grav.     & H$_2$O-1   & H$_2$O-2   &  FeH  \\
Value & 1.17     & 1.01      & 0.660      & 0.831      & 0.877  \\
SpT   & L0       & low grav. & L0         & L1.5       & M8.5   \\
\hline      
Index & H$_2$O-A   & H$_2$O-B   & H$_2$O-C   & H$_2$O-D   & \\
Value & 0.593      & 0.724      & 0.733      & 0.817      & \\
SpT   & L2.5       & L3         & L0         & L4         & \\
\hline                  

\end{tabular}
\end{minipage}
\end{table}

As noted previously, to match the predicted spectral model fluxes to the
observed photometry, either 2M1207\,b is underluminous or has a smaller than
expected radius. The required radius to explain the photometry is
$\sim$0.052\,$R_{\sun}$, about one third of the radius of
$\sim$0.16\,$R_{\sun}$ from the DUSTY evolutionary model prediction for an
object of \teff=1600\,K at 5--10\,Myr, and significantly smaller than any
predicted radius for a partially degenerate object above 1\mjup\ \citep{c09}.

The origin of 2M1207\,b and its underluminous flux is uncertain, and
different scenarios have been proposed for the specific case of 2M1207\,b: an
edge-on disk  \citep{m07} and the result of a collision of protoplanets
\citep{mam07}. More generally, the impact on the luminosity due to initial
conditions and the early accretion history  of young objects with ages
comparable to 2M1207\,b, has been investigated \citep{mar07, b09}. The
protoplanet collision hypothesis predicts that the object is much smaller and
should have a surface gravity of 3.0 rather than 4.5. The upper panel of
Fig.\,\ref{2M1207b-spectra} shows the data with DUSTY spectral models with
three different effective temperatures for a lower surface gravity of
\logg=3.5 (since \logg=3.0 is not available). The data favour the higher
surface gravity \logg=4.5 value, though increasing the $J$-band flux by one
sigma would make the data consistent with the \logg=3.5 model. A $J$-band
magnitude difference with smaller uncertainties would be required to entirely
rule out the collision scenario. 

For an initial investigation into the edge-on disk possibility, we compared
the photometry and spectra with synthetic spectra generated by the TORUS and
MCFOST codes \citep{p09}. A central object simulated by the \teff=1600 K DUSTY
atmosphere was encircled with a disk 10 AU in radius, consistent with the
physical separation of the 2M1207 pair. The disk was constructed with canonical values for
flaring, surface density power law, dust size distribution and chemistry. The
inclination was allowed to vary from edge-on to the $\sim$60 degree limit
suggested for the geometry of the primary disk. Within the limited range of
disk parameters searched, scattering alone could not explain the flux level of
the central object since the predicted scattered flux was only a few per-cent,
rather than the required $\sim$10\%. For comparison, the T Tauri stars HH 30
and HK Tau with imaged edge-on disks \citep{s98} are fainter by 3--4
magnitudes relative to the photospheric levels, consistent with a
scattered flux level of a few per-cent or less. 
Thus, among disk scenarios, grey extinction from a disk composed of larger
grains than considered in the simulations is favoured over scattering off the
surface of an optically thick disk; grain sizes larger than 4\micron\ have
been suggested as a possibility \citep{m07}. Additionally, extinction of a
hotter object is not able to explain the wavelength dependence of the
photometry over the entire 0.9--3.6\micron\ range.

In summary, the SINFONI spectrum of 2M1207\,b represents the highest
resolution, largest wavelength coverage spectrum of a young, planetary mass
companion.  The overall shape is distinct from field and young cluster
L-dwarfs, with substantially different ratios of flux across the bands, a
triangular shape to the $H$-band continuum, and a peak at longer wavelengths
in the $K$-band. Spectral indices developed for cool objects give conflicting
values, indicating the unusual nature of this object.  The spectrum is
best fit by a DUSTY model atmosphere with \teff=1600\,K and \logg= 4.5,
though it is not yet possible to exclude the \logg= 3.0 atmosphere
predicted by one formation model. We find that the apparent underluminosity
of 2M1207\,b is consistent with extinction from large dust grains in a nearly
edge-on disk, but more precise $J$-band and longer wavelength photometry and
spectroscopy will help to determine the nature of this important object. This
spectrum of 2M1207\,b will serve as a key comparison for upcoming
near-infrared spectroscopy of planets imaged around stellar hosts. 

\begin{acknowledgements} 

      We gratefully acknowledge funding for a  Leverhulme research  project
      grant to J. P. (ID20090041) and a studenship for R. J. DR. from the
      Science and Technology Facilities Council (STFC). We thank T. Harries,
      C. Pinte, and N. Mayne for providing synthetic SEDs for comparison with
      the data. We thank J. Fortney, R. White and I. Baraffe for 
      scientific discussions and the referee for helpful comments. 

\end{acknowledgements}

\bibliographystyle{aa}
\bibliography{bib}

\begin{thebibliography}{44}
\expandafter\ifx\csname natexlab\endcsname\relax\def\natexlab#1{#1}\fi

\bibitem[{{Allard} {et~al.}(2001){Allard}, {Hauschildt}, {Alexander},
  {Tamanai}, \& {Schweitzer}}]{a01}
{Allard}, F., {Hauschildt}, P.~H., {Alexander}, D.~R., {Tamanai}, A., \&
  {Schweitzer}, A. 2001, \apj, 556, 357

\bibitem[{{Allers} {et~al.}(2007){Allers}, {Jaffe}, {Luhman}, {Liu}, {Wilson},
  {Skrutskie}, {Nelson}, {Peterson}, {Smith}, \& {Cushing}}]{a07}
{Allers}, K.~N., {Jaffe}, D.~T., {Luhman}, K.~L., {et~al.} 2007, \apj, 657, 511

\bibitem[{{Baraffe} {et~al.}(2002){Baraffe}, {Chabrier}, {Allard}, \&
  {Hauschildt}}]{b02}
{Baraffe}, I., {Chabrier}, G., {Allard}, F., \& {Hauschildt}, P.~H. 2002, \aap,
  382, 563

\bibitem[{{Baraffe} {et~al.}(2003){Baraffe}, {Chabrier}, {Barman}, {Allard}, \&
  {Hauschildt}}]{b03}
{Baraffe}, I., {Chabrier}, G., {Barman}, T.~S., {Allard}, F., \& {Hauschildt},
  P.~H. 2003, \aap, 402, 701

\bibitem[{{Baraffe} {et~al.}(2009){Baraffe}, {Chabrier}, \& {Gallardo}}]{b09}
{Baraffe}, I., {Chabrier}, G., \& {Gallardo}, J. 2009, \apjl, 702, L27

\bibitem[{{Bonnefoy} {et~al.}(2010){Bonnefoy}, {Chauvin}, {Rojo}, {Allard},
  {Lagrange}, {Homeier}, {Dumas}, \& {Beuzit}}]{b10}
{Bonnefoy}, M., {Chauvin}, G., {Rojo}, P., {et~al.} 2010, ArXiv e-prints

\bibitem[{{Bonnet} {et~al.}(2004){Bonnet}, {Abuter}, {Baker}, {Bornemann},
  {Brown}, {Castillo}, {Conzelmann}, {Damster}, {Davies}, {Delabre},
  {Donaldson}, {Dumas}, {Eisenhauer}, {Elswijk}, {Fedrigo}, {Finger},
  {Gemperlein}, {Genzel}, {Gilbert}, {Gillet}, {Goldbrunner}, {Horrobin}, {Ter
  Horst}, {Huber}, {Hubin}, {Iserlohe}, {Kaufer}, {Kissler-Patig}, {Kragt},
  {Kroes}, {Lehnert}, {Lieb}, {Liske}, {Lizon}, {Lutz}, {Modigliani}, {Monnet},
  {Nesvadba}, {Patig}, {Pragt}, {Reunanen}, {R{\"o}hrle}, {Rossi}, {Schmutzer},
  {Schoenmaker}, {Schreiber}, {Stroebele}, {Szeifert}, {Tacconi}, {Tecza},
  {Thatte}, {Tordo}, {van der Werf}, \& {Weisz}}]{b04}
{Bonnet}, H., {Abuter}, R., {Baker}, A., {et~al.} 2004, The Messenger, 117, 17

\bibitem[{{Boss}(1997)}]{b97}
{Boss}, A.~P. 1997, Science, 276, 1836

\bibitem[{{Carpenter}(2001)}]{Carpenter:2001}
{Carpenter}, J.~M. 2001, \aj, 121, 2851

\bibitem[{{Chabrier} {et~al.}(2000){Chabrier}, {Baraffe}, {Allard}, \&
  {Hauschildt}}]{c00}
{Chabrier}, G., {Baraffe}, I., {Allard}, F., \& {Hauschildt}, P. 2000, \apj,
  542, 464

\bibitem[{{Chabrier} {et~al.}(2009){Chabrier}, {Baraffe}, {Leconte},
  {Gallardo}, \& {Barman}}]{c09}
{Chabrier}, G., {Baraffe}, I., {Leconte}, J., {Gallardo}, J., \& {Barman}, T.
  2009, in American Institute of Physics Conference Series, Vol. 1094, American
  Institute of Physics Conference Series, ed. {E.~Stempels}, 102--111

\bibitem[{{Charbonneau} {et~al.}(2005){Charbonneau}, {Allen}, {Megeath},
  {Torres}, {Alonso}, {Brown}, {Gilliland}, {Latham}, {Mandushev}, {O'Donovan},
  \& {Sozzetti}}]{ch05}
{Charbonneau}, D., {Allen}, L.~E., {Megeath}, S.~T., {et~al.} 2005, \apj, 626,
  523

\bibitem[{{Charbonneau} {et~al.}(2002){Charbonneau}, {Brown}, {Noyes}, \&
  {Gilliland}}]{c02}
{Charbonneau}, D., {Brown}, T.~M., {Noyes}, R.~W., \& {Gilliland}, R.~L. 2002,
  \apj, 568, 377

\bibitem[{{Chauvin} {et~al.}(2004){Chauvin}, {Lagrange}, {Dumas}, {Zuckerman},
  {Mouillet}, {Song}, {Beuzit}, \& {Lowrance}}]{c04}
{Chauvin}, G., {Lagrange}, A., {Dumas}, C., {et~al.} 2004, \aap, 425, L29

\bibitem[{{Chauvin} {et~al.}(2005){Chauvin}, {Lagrange}, {Zuckerman}, {Dumas},
  {Mouillet}, {Song}, {Beuzit}, {Lowrance}, \& {Bessell}}]{c05}
{Chauvin}, G., {Lagrange}, A., {Zuckerman}, B., {et~al.} 2005, \aap, 438, L29

\bibitem[{{Deming} {et~al.}(2005){Deming}, {Seager}, {Richardson}, \&
  {Harrington}}]{d05}
{Deming}, D., {Seager}, S., {Richardson}, L.~J., \& {Harrington}, J. 2005,
  \nat, 434, 740

\bibitem[{{Ducourant} {et~al.}(2008){Ducourant}, {Teixeira}, {Chauvin},
  {Daigne}, {Le Campion}, {Song}, \& {Zuckerman}}]{d08}
{Ducourant}, C., {Teixeira}, R., {Chauvin}, G., {et~al.} 2008, \aap, 477, L1

\bibitem[{{Dumas} {et~al.}(2007){Dumas}, {Kaufer}, \& {Hainaut}}]{d07}
{Dumas}, C., {Kaufer}, A., \& {Hainaut}, O. 2007, in SINFONI data reduction
  cookbook, VLT-MAN-ESO-14700-4037

\bibitem[{{Eisenhauer} {et~al.}(2003){Eisenhauer}, {Abuter}, {Bickert},
  {Biancat-Marchet}, {Bonnet}, {Brynnel}, {Conzelmann}, {Delabre}, {Donaldson},
  {Farinato}, {Fedrigo}, {Genzel}, {Hubin}, {Iserlohe}, {Kasper},
  {Kissler-Patig}, {Monnet}, {Roehrle}, {Schreiber}, {Stroebele}, {Tecza},
  {Thatte}, \& {Weisz}}]{e03}
{Eisenhauer}, F., {Abuter}, R., {Bickert}, K., {et~al.} 2003, in SPIE
  Conference Series, Vol. 4841, , 1548--1561

\bibitem[{{Fortney} {et~al.}(2008){Fortney}, {Marley}, {Saumon}, \&
  {Lodders}}]{f08}
{Fortney}, J.~J., {Marley}, M.~S., {Saumon}, D., \& {Lodders}, K. 2008, \apj,
  683, 1104

\bibitem[{{Gebbinck} {et~al.}(2007){Gebbinck}, {Ballester}, \& {Peron}}]{g07}
{Gebbinck}, M.~K., {Ballester}, P., \& {Peron}, M. 2007, in VLT Gasgano User's
  Manual, VLT-PRO-ESO-19000-1932

\bibitem[{{Gizis}(2002)}]{g02}
{Gizis}, J.~E. 2002, \apj, 575, 484

\bibitem[{{Grillmair} {et~al.}(2008){Grillmair}, {Burrows}, {Charbonneau},
  {Armus}, {Stauffer}, {Meadows}, {van Cleve}, {von Braun}, \& {Levine}}]{g08}
{Grillmair}, C.~J., {Burrows}, A., {Charbonneau}, D., {et~al.} 2008, \nat, 456,
  767

\bibitem[{{Hayes}(1985)}]{Hayes:1985}
{Hayes}, D.~S. 1985, in IAU Symposium, Vol. 111, Calibration of Fundamental
  Stellar Quantities, ed. D.~S. {Hayes}, L.~E. {Pasinetti}, \& A.~G.~D.
  {Philip}, 225--249

\bibitem[{{Janson} {et~al.}(2010){Janson}, {Bergfors}, {Goto}, {Brandner}, \&
  {Lafreni{\`e}re}}]{j10}
{Janson}, M., {Bergfors}, C., {Goto}, M., {Brandner}, W., \& {Lafreni{\`e}re},
  D. 2010, \apjl, 710, L35

\bibitem[{{Kalas} {et~al.}(2008){Kalas}, {Graham}, {Chiang}, {Fitzgerald},
  {Clampin}, {Kite}, {Stapelfeldt}, {Marois}, \& {Krist}}]{k08}
{Kalas}, P., {Graham}, J.~R., {Chiang}, E., {et~al.} 2008, Science, 322, 1345

\bibitem[{{Lodieu} {et~al.}(2008){Lodieu}, {Hambly}, {Jameson}, \&
  {Hodgkin}}]{l08}
{Lodieu}, N., {Hambly}, N.~C., {Jameson}, R.~F., \& {Hodgkin}, S.~T. 2008,
  \mnras, 383, 1385

\bibitem[{{Mamajek} \& {Meyer}(2007)}]{mam07}
{Mamajek}, E.~E. \& {Meyer}, M.~R. 2007, \apjl, 668, L175

\bibitem[{{Marley} {et~al.}(2007){Marley}, {Fortney}, {Hubickyj},
  {Bodenheimer}, \& {Lissauer}}]{mar07}
{Marley}, M.~S., {Fortney}, J.~J., {Hubickyj}, O., {Bodenheimer}, P., \&
  {Lissauer}, J.~J. 2007, \apj, 655, 541

\bibitem[{{Marois} {et~al.}(2008){Marois}, {Macintosh}, {Barman}, {Zuckerman},
  {Song}, {Patience}, {Lafreni{\`e}re}, \& {Doyon}}]{m08}
{Marois}, C., {Macintosh}, B., {Barman}, T., {et~al.} 2008, Science, 322, 1348

\bibitem[{{McLean} {et~al.}(2003){McLean}, {McGovern}, {Burgasser},
  {Kirkpatrick}, {Prato}, \& {Kim}}]{m03}
{McLean}, I.~S., {McGovern}, M.~R., {Burgasser}, A.~J., {et~al.} 2003, \apj,
  596, 561

\bibitem[{{Modigliani} {et~al.}(2009){Modigliani}, {Mirny}, {Ballester}, \&
  {Peron}}]{m09}
{Modigliani}, A., {Mirny}, K., {Ballester}, P., \& {Peron}, M. 2009, in SINFONI
  Pipeline User Manual, VLT-MAN-ESO-19500-3600

\bibitem[{{Mohanty} {et~al.}(2007){Mohanty}, {Jayawardhana}, {Hu{\'e}lamo}, \&
  {Mamajek}}]{m07}
{Mohanty}, S., {Jayawardhana}, R., {Hu{\'e}lamo}, N., \& {Mamajek}, E. 2007,
  \apj, 657, 1064

\bibitem[{{Mountain} {et~al.}(1985){Mountain}, {Selby}, {Leggett}, {Blackwell},
  \& {Petford}}]{Mountain:1985}
{Mountain}, C.~M., {Selby}, M.~J., {Leggett}, S.~K., {Blackwell}, D.~E., \&
  {Petford}, A.~D. 1985, \aap, 151, 399

\bibitem[{{Pinte} {et~al.}(2009){Pinte}, {Harries}, {Min}, {Watson},
  {Dullemond}, {Woitke}, {M{\'e}nard}, \& {Dur{\'a}n-Rojas}}]{p09}
{Pinte}, C., {Harries}, T.~J., {Min}, M., {et~al.} 2009, \aap, 498, 967

\bibitem[{{Pollack} {et~al.}(1996){Pollack}, {Hubickyj}, {Bodenheimer},
  {Lissauer}, {Podolak}, \& {Greenzweig}}]{p96}
{Pollack}, J.~B., {Hubickyj}, O., {Bodenheimer}, P., {et~al.} 1996, Icarus,
  124, 62

\bibitem[{{Pont} {et~al.}(2008){Pont}, {Knutson}, {Gilliland}, {Moutou}, \&
  {Charbonneau}}]{p08}
{Pont}, F., {Knutson}, H., {Gilliland}, R.~L., {Moutou}, C., \& {Charbonneau},
  D. 2008, \mnras, 385, 109

\bibitem[{{Rhee} {et~al.}(2007){Rhee}, {Song}, {Zuckerman}, \&
  {McElwain}}]{r07}
{Rhee}, J.~H., {Song}, I., {Zuckerman}, B., \& {McElwain}, M. 2007, \apj, 660,
  1556

\bibitem[{{Sing} {et~al.}(2008){Sing}, {Vidal-Madjar}, {Lecavelier des Etangs},
  {D{\'e}sert}, {Ballester}, \& {Ehrenreich}}]{s08}
{Sing}, D.~K., {Vidal-Madjar}, A., {Lecavelier des Etangs}, A., {et~al.} 2008,
  \apj, 686, 667

\bibitem[{{Slesnick} {et~al.}(2004){Slesnick}, {Hillenbrand}, \&
  {Carpenter}}]{s04}
{Slesnick}, C.~L., {Hillenbrand}, L.~A., \& {Carpenter}, J.~M. 2004, \apj, 610,
  1045

\bibitem[{{Song} {et~al.}(2006){Song}, {Schneider}, {Zuckerman}, {Farihi},
  {Becklin}, {Bessell}, {Lowrance}, \& {Macintosh}}]{s06}
{Song}, I., {Schneider}, G., {Zuckerman}, B., {et~al.} 2006, \apj, 652, 724

\bibitem[{{Stapelfeldt} {et~al.}(1998){Stapelfeldt}, {Krist}, {Menard},
  {Bouvier}, {Padgett}, \& {Burrows}}]{s98}
{Stapelfeldt}, K.~R., {Krist}, J.~E., {Menard}, F., {et~al.} 1998, \apjl, 502,
  L65+

\bibitem[{{Swain} {et~al.}(2008){Swain}, {Vasisht}, \& {Tinetti}}]{sw08}
{Swain}, M.~R., {Vasisht}, G., \& {Tinetti}, G. 2008, \nat, 452, 329

\bibitem[{{Thatte} {et~al.}(1998){Thatte}, {Tecza}, {Eisenhauer}, {Mengel},
  {Krabbe}, {Pak}, {Genzel}, {Bonaccini}, {Emsellem}, {Rigaut}, {Delabre}, \&
  {Monnet}}]{t98}
{Thatte}, N.~A., {Tecza}, M., {Eisenhauer}, F., {et~al.} 1998, in SPIE
  Conference Series, ed. {D.~Bonaccini \& R.~K.~Tyson}, Vol. 3353, 704--715

\end{thebibliography}

\end{document}